\documentclass[12pt]{article}
\usepackage{amsfonts}
\usepackage{bm}
\usepackage{graphicx}
\usepackage{multicol}

\begin{document}

\title{\emph{}
Features of free particles system motion in noncommutative phase space
}
\maketitle

\centerline {Kh. P. Gnatenko \footnote{E-Mail address: khrystyna.gnatenko@gmail.com}, H. P. Laba \footnote{E-Mail address: hanna.laba@polynet.lviv.ua}, V. M. Tkachuk \footnote{E-Mail address: voltkachuk@gmail.com}}
\medskip

\centerline {\small  $^{1,3}$ \it  Ivan Franko National University of Lviv, }
\centerline {\small \it Department for Theoretical Physics,12 Drahomanov St., Lviv, 79005, Ukraine}

\centerline {\small  $^{2}$ \it  Lviv Polytechnic National University,}
\centerline {\small \it  Department of Applied Physics and Nanomaterials Science, 5 Ustiyanovych St., Lviv, 79013, Ukraine}
\begin{abstract}
 Influence of noncommutativity on the motion of composite system is studied in noncommutative phase space of canonical type. A system composed by $N$ free particles is examined. We show that because of momentum noncommutativity free particles of different masses with the same velocities at the initial moment of time do not move together. The trajectory and the velocity of free particle in noncommutative phase space depend on its mass. So, a system of the free particles flies away.  Also, it is shown that the total momentum defined in the traditional way is not integral of motion in a space with noncommutativity of coordinates and noncommutativity of momenta. We find that in the case when parameters of noncommutativity corresponding to a particle are determined by its mass  the trajectory and velocity of free particle are independent of the mass, also the total momenta as integrals of motion can be introduced in noncommutative phase space.


 PACS number(s): 11.90.+t, 11.10.Nx

\end{abstract}
\section{Introduction}

Studies of quantum space realized on the basis of idea of noncommutativity of coordinates  have obtained grate interest recently. The interest is due to development of String theory and Quantum Gravity (see, for instance, \cite{Witten,Doplicher}). The idea of noncommutativity was proposed by Heisenberg, later it was formalized  by Snyder and published in his paper \cite{Snyder}.

Much attention has been devoted to studies of physical systems in quantum space realized on the basis of idea of noncommutativity. Among them are studies of  harmonic oscillator \cite{Hatzinikitas,Kijanka,Jing,Smailagic,Smailagic1,Alvarez,Djemai,Giri,Geloun,Nath,GnatenkoJPS17,Shyiko}, Landau problem  \cite{Gamboa1,Horvathy,Dayi,Horvathy05,Alvarez09,Daszkiewicz1},  particle in gravitational field \cite{Bertolami1,Bastos,GnatenkoMPLA16,GnatenkoJPS18}, free quantum particle \cite{Djemai,Shyiko}, quantum fields  \cite{Balachandran10,Balachandran11} and many others.

 In four dimensional  case (2D configurational space and 2D momentum space) the coordinates and the momenta in the space satisfy the following relations
 \begin{eqnarray}
[X_{1},X_{2}]=i\hbar\theta,\label{form101}\\{}
[X_{i},P_{j}]=i\hbar\delta_{ij},\label{form1001}\\{}
[P_{1},P_{2}]=i\hbar\eta.\label{form10001}{}
\end{eqnarray}
 where $\theta=const$ is parameter of coordinate noncommutativity, $\eta=const$ is parameter of momentum noncommutativity, $i,j=(1,2)$. In the classical limit from  (\ref{form101})-(\ref{form10001}) one obtains deformed Poisson brackets
\begin{eqnarray}
\{X_{1},X_{2}\}=\theta,\label{al00}\\{}
\{X_{i},P_{j}\}=\delta_{ij},\\{}
\{P_{1},P_{2}\}=\eta.\label{al10}
\end{eqnarray}

 Studies of many particle problems in quantum space open new possibilities to find effects of space quantitization on the Planck scale in properties of a wide class of physical systems. In a space with noncommutativity of coordinates the problems of quantum mechanics of many particles were examined in \cite{Ho}.  Authors of paper \cite{Bellucci} studied the system of two charged quantum particles in noncommutative space. Features of  motion of composite system in gravitational field in a space with coordinates noncommutativity were considered in \cite{GnatenkoPLA13,GnatenkoJPS13}. Two-particle system interacting through the harmonic oscillator potential on a noncommutative plane was examined in \cite{Jabbari}. Spectrum of two-particle system with Coulomb interaction was studied in rotationally-invariant space with noncommutativity of coordinates in \cite{GnatenkoU16}. Also a system of two particles was studied in quantum space characterized by coordinate noncommutativity and momentum noncommutativity \cite{Djemai}.
In noncommutative space-time the classical problem of many particles was examined in \cite{Daszkiewicz}. In \cite{Daszkiewicz2} the quantum model of many particles moving in twisted N-enlarged Newton-Hooke space-time was proposed. In \cite{GnatenkoPLA17} the properties of kinetic energy of composite system and its motion in the gravitational field were studied in four-dimensional noncommutative phase space. Features in description of composite system in rotationally-invariant noncommutative phase space were considered in \cite{GnatenkoIJ18}.

The problem of composite system which is known as soccer-ball
problem was considered in the case of Doubly Special Relativity \cite{Jacobson,Amelino-Camelia} relative locality   \cite{Amelino-Camelia1,Hossenfelder,Amelino-Camelia3}.

In the deformed space with minimal length the problem of description of motion of composite system was studied in \cite{Quesne}. The authors of \cite{Quesne} introduced the total momenta as integrals of motion. The motion of composite system in gravitational field in the space was examined in \cite{Tkachuk}.

In the present paper we study features of motion of free particles  system in noncommutative phase space of canonical type in the general case when different particles satisfy noncommutative algebra with different parameters of noncommutativity. We show that even in the case when at the initial moment of time the velocities of particles in the system are the same  the particles fly away. The situation is changed if we consider parameters of noncommutativity which correspond to a particle to be dependent on its mass as was proposed in \cite{GnatenkoPLA17,GnatenkoMPLA17}. In noncommutative phase space the total momentum defined in the traditional way (defined as a sum of momenta of particles) is not preserved. In the present paper we find total momenta of composite system in noncommutative phase space which are integrals of motion and introduce the coordinates of the center-of-mass conjugated to them.

The paper is organized as follows. In Section 2 the motion of composite system in noncommutative phase space is considered. In Section 3 features of motion of free particles system in noncommutative phase space are examined. The total momenta as integrals of motion and coordinates of the center-of-mass conjugated to them are introduced in Section 4. Conclusions are presented in Section 5.

\section{Description of motion of composite system in noncommutative phase space}

Let us consider composite system made of $N$ particles with masses $m_a$ $(a=1..N)$  with the following hamiltonian
\begin{eqnarray}
  H=\sum_{a}\frac{({\bf P}^{(a)})^{2}}{2m_{a}}+\frac{1}{2}\mathop{\sum_{a,b}}\limits_{a\neq b} U(|{\bf X}^{(a)}-{\bf X}^{(b)}|),\label{hham}
\end{eqnarray}
in four-dimensional noncommutative phase space. In (\ref{hham}) indexes $a$, $b$ label the particles.

In general case coordinates and momenta of different particles may satisfy noncommutative algebra with different parameters of noncommutativity
\begin{eqnarray}
[X_{1}^{(a)},X_{2}^{(b)}]=i\hbar\delta^{ab}\theta_{a},\label{al0}\\{}
[X_{i}^{(a)},P_{j}^{(b)}]=i\hbar\delta^{ab}\delta_{ij},\\{}
[P_{1}^{(a)},P_{2}^{(b)}]=i\hbar\delta^{ab}\eta_{a},\label{al1}
\end{eqnarray}
where  $\theta_{a}$, $\eta_{a}$ are parameters of noncommutativity which correspond to particle with mass $m_a$. In the classical limit we have the following Poisson brackets
\begin{eqnarray}
\{X_{1}^{(a)},X_{2}^{(b)}\}=\delta^{ab}\theta_{a},\label{pal0}\\{}
\{X_{i}^{(a)},P_{j}^{(b)}\}=\delta^{ab}\delta_{ij},\\{}
\{P_{1}^{(a)},P_{2}^{(b)}\}=\delta^{ab}\eta_{a}.\label{pal1}
\end{eqnarray}

In our previous paper \cite{GnatenkoPLA17} we show that the coordinates and momenta of the center-of-mass, the  coordinates and momenta of the relative motion which are defined in the traditional way
 \begin{eqnarray}
\tilde{{\bf P}}=\sum_{a}{\bf P}^{(a)},\label{05}\\
\tilde{{\bf X}}=\sum_{a}\mu_{a}{\bf X}^{(a)},\label{00005}\\
\Delta{\bf P}^{{a}}={\bf P}^{(a)}-\mu_{a}\tilde{{\bf P}},\\
{\Delta\bf X}^{(a)}={\bf X}^{(a)}-\tilde{{\bf X}},\label{06}
\end{eqnarray}
with $\mu_a=m_{a}/M$, $M=\sum_{a}m_a$, satisfy noncommutative algebra with effective parameters of noncommutativity
\begin{eqnarray}
\tilde{\theta}=\frac{\sum_{a}m_{a}^{2}\theta_{a}}{(\sum_{b}m_{b})^{2}},\label{eff}\\
\tilde{\eta}=\sum_{a}\eta_a.\label{eff2}
\end{eqnarray}
The Poisson brackets for $\tilde{X}_i$, $\tilde{P}_i$, $\Delta{X}_i^{(a)}$, $\Delta{P}_i^{(a)}$ read
\begin{eqnarray}
\{\tilde{X}_1,\tilde{X}_2\}=\tilde{\theta},\label{07}\\{}
\{\tilde{P}_1,\tilde{P}_2\}=\tilde{\eta},\\{}
\{\tilde{X}_i,\tilde{P}_j\}=\{\Delta{X}_i,\Delta{P}_j\}=\delta_{ij},\label{08}\\{}
\{\Delta{X}_1^{(a)},\Delta{X}_2^{(b)}\}=-\{\Delta{X}_2^{(a)},\Delta{X}_1^{(b)}\}=\delta^{ab}\theta_{a}-\mu_{a}\theta_{a}-\mu_{b}\theta_{b}+\tilde{\theta} ,\\{}
\{\Delta{P}_1^{(a)},\Delta{P}_2^{(b)}\}=-\{\Delta{P}_2^{(a)},\Delta{P}_1^{(b)}\}=\delta^{ab}\eta_a-\mu_b\eta_a-\mu_a\eta_b+\mu_a\mu_b\tilde{\eta}.{}\label{007}
\end{eqnarray}
Also, in the paper \cite{GnatenkoPLA17} we mentioned that the Poisson brackets for coordinates of the center-of-mass and coordinates of the relative motion and Poisson brackets  for the total momenta and the momenta of relative motion do not equal to zero. Namely, the following relations are satisfied
 \begin{eqnarray}
\{\tilde{X}_{1},\Delta X_{2}^{(a)}\}=-\{\tilde{X}_{2},\Delta X_{1}^{(a)}\}=\mu_{a}\theta_{a}-\tilde{\theta},\label{rel1}\\{}
\{\tilde{P}_1,\Delta{P}^{(a)}_2\}=-\{\tilde{P}_2,\Delta{P}^{(a)}_1\}=\eta_a-\mu_a\sum_{b}\eta_b.\label{rel2}
\end{eqnarray}
So, in noncommutative phase space we cannot consider hamiltonian of the center-of-mass and hamiltonian of the relative motion as independent. From (\ref{hham}), using (\ref{05})-(\ref{06}), we have
 \begin{eqnarray}
H=H_{cm}+H_{rel},\\
H_{cm}=\frac{\tilde{{\bf P}}^{2}}{2M},\label{12cm}\\
H_{rel}=\sum_a\frac{(\Delta{\bf P}^{a})^{2}}{2m_a}+\frac{1}{2}\mathop{\sum_{a,b}}\limits_{a\neq b} U(|\Delta{\bf X}^{(a)}-\Delta{\bf X}^{(b)}|)
 \end{eqnarray}
and $\{H_{cm},H_{rel}\}\neq0$.
In paper \cite{GnatenkoPLA17} we proposed conditions on the parameters of noncommutativity
\begin{eqnarray}
\frac{\eta_a}{m_a}=\alpha=const,\label{cond}\\
\theta_a m_a=\gamma=const,\label{cond2}
\end{eqnarray}
with $\alpha$, $\gamma$ being constants which are the same for particles with different masses.  On these conditions $\{\tilde{X}_{1},\Delta X_{2}^{(a)}\}=0$, $\{\tilde{P}_1,\Delta{P}^{a}_2\}=0$, therefore the motion of the center-of-mass is independent of the relative motion.

\section{Motion of free particles system and parameters of noncommutativity}

Let us study  features of motion of a free particle of mass $m$ in four-dimensional noncommutative phase space. The hamiltonian of the particle  reads
\begin{eqnarray}
H=\frac{{P}_1^2}{2m}+\frac{{P}_2^2}{2m}.\label{hammm}
\end{eqnarray}
Taking into account (\ref{al00})-(\ref{al10}), (\ref{hammm}) we can write the following equations of motion
\begin{eqnarray}
\dot{X}_1=\frac{P_1}{m},\label{eq}\\
\dot{X}_2=\frac{P_2}{m},\label{eq0}\\
\dot{P}_1=\eta\frac{P_2}{m},\label{eq1111}\\
\dot{P}_2=-\eta\frac{P_1}{m}.\label{eq1}
\end{eqnarray}
Solutions of equations (\ref{eq})-(\ref{eq1}) with initial conditions ${X}_1(0)=X_{01}$, ${X}_2(0)=X_{02}$, $\dot{X}_1(0)=\upsilon_{01}$ $\dot{X}_2(0)=\upsilon_{02}$  are as follows
\begin{eqnarray}
{X}_1(t)=\upsilon_{01}\frac{m}{\eta}\sin\frac{\eta}{m}t-\upsilon_{02}\frac{m}{\eta}\cos\frac{\eta}{m}t+\upsilon_{02}\frac{m}{\eta}+X_{01},\label{eqs}\\
{X}_2(t)=\upsilon_{02}\frac{m}{\eta}\sin\frac{\eta}{m}t+\upsilon_{01}\frac{m}{\eta}\cos\frac{\eta}{m}t-\upsilon_{01}\frac{m}{\eta}+X_{02},\label{eqs22}\\
\dot{X}_1(t)=\upsilon_{01}\cos\frac{\eta}{m}t+\upsilon_{02}\sin\frac{\eta}{m}t,\label{eqsd}\\
\dot{X}_2(t)=\upsilon_{02}\cos\frac{\eta}{m}t-\upsilon_{01}\sin\frac{\eta}{m}t.\label{eqsd1}
\end{eqnarray}
In the limit $\theta\rightarrow0$, $\eta\rightarrow0$ we obtain trajectory of the particle in the ordinary space ${X}_1(t)=\upsilon_{01}t+X_{01}$, ${X}_2(t)=\upsilon_{02}t+X_{02}$.
Note that because of noncommutativity of momenta the trajectory and the velocity of free particle depend on its mass and on the parameter $\eta$.

Let us consider a system of free particles with masses $m_1$, $m_2$,...,$m_N$ with hamiltonian
 \begin{eqnarray}
H=\sum_a\frac{({\bf P}^{(a)})^2}{2m_a}=\frac{\tilde{{\bf P}}^2}{2M}+\sum_a\frac{(\Delta{\bf P}^{(a)})^2}{2m_a}.\label{hsys}
\end{eqnarray}
The coordinates $X_i^{(a)}$ and momenta $P_i^{(a)}$ of particles satisfy relations (\ref{pal0})-(\ref{pal1}). Momenta of the center-of-mass $\tilde P_i$ and  momenta of relative motion $\Delta P_i^{(a)}$ satisfy  (\ref{07})-(\ref{rel2}), $M=\sum_a m_a$.

In the ordinary space ($\theta=0$, $\eta=0$) free particles with the same initial velocities move together.
In noncommutative phase space even in the case of equality of velocities of free particles of masses $m_1$, $m_2$,...,$m_N$ at the initial moment of time $\dot{X}^{(a)}_1(0)=\upsilon_{01}$, $\dot{X}^{(a)}_2(0)=\upsilon_{02}$, $a=(1...N)$, we have $\dot{X}^{(a)}_1(t)\neq\dot{X}^{(b)}_1(t)$, $\dot{X}^{(a)}_2(t)\neq\dot{X}^{(b)}_2(t)$ for $a\neq b$. Namely using (\ref{eqsd}), (\ref{eqsd1}), one can write
\begin{eqnarray}
\dot{X}^{(a)}_1(t)=\upsilon_{01}\cos\frac{\eta_a}{m_a}t+\upsilon_{02}\sin\frac{\eta_a}{m_a}t,\label{eqsda}\\
\dot{X}^{(a)}_2(t)=\upsilon_{02}\cos\frac{\eta_a}{m_a}t-\upsilon_{01}\sin\frac{\eta_a}{m_a}t,\label{eqsd1a}
\end{eqnarray}
here $\eta_a$ is parameter of momentum noncommutativity which corresponds to particle of mass $m_a$, $a=(1..N)$.

Note that even for a system of free particles the relative motion affects on the motion of the center-of-mass because of relation  (\ref{rel2}). The trajectory of the center-of-mass of the system and the trajectory of relative motion read
\begin{eqnarray}
\tilde{X}_1(t)=\sum_a\left(\upsilon_{01}\frac{m^2_a}{M\eta_a}\sin\frac{\eta_a}{m_a}t-\upsilon_{02}\frac{m^2_a}{M\eta_a}\cos\frac{\eta_a}{m_a}t+\upsilon_{02}\frac{m^2_a}{M\eta_a}+\frac{m_a}{M}X^{(a)}_{01}\right),\label{peqs}\\
\tilde{X}_2(t)=\sum_a\left(\upsilon_{02}\frac{m^2_a}{M\eta_a}\sin\frac{\eta_a}{m_a}t+\upsilon_{01}\frac{m^2_a}{M\eta_a}\cos\frac{\eta_a}{m_a}t-\upsilon_{01}\frac{m^2_a}{M\eta_a}+\frac{m_a}{M}X^{(a)}_{02}\right),\label{peqs22}\\
\Delta{X}^a_1(t)=\upsilon_{01}\frac{m_a}{\eta_a}\sin\frac{\eta_a}{m_a}t-\upsilon_{02}\frac{m_a}{\eta_a}\cos\frac{\eta_a}{m_a}t+\upsilon_{02}\frac{m_a}{\eta_a}+X^{(a)}_{01}-\nonumber\\
-\sum_b\left(\upsilon_{01}\frac{m^2_b}{M\eta_b}\sin\frac{\eta_b}{m_b}t-\upsilon_{02}\frac{m^2_b}{M\eta_b}\cos\frac{\eta_b}{m_b}t+\upsilon_{02}\frac{m^2_b}{M\eta_b}+\frac{m_b}{M}X^{(b)}_{01}\right),\label{21}\\
\Delta{X}^a_2(t)=\upsilon_{02}\frac{m_a}{\eta_a}\sin\frac{\eta_a}{m_a}t+\upsilon_{01}\frac{m_a}{\eta_a}\cos\frac{\eta_a}{m_a}t-\upsilon_{01}\frac{m_a}{\eta_a}+X^{(a)}_{02}-\nonumber\\
-\sum_b\left(\upsilon_{02}\frac{m^2_b}{M\eta_b}\sin\frac{\eta_b}{m_b}t+\upsilon_{01}\frac{m^2_b}{M\eta_b}\cos\frac{\eta_b}{m_b}t-\upsilon_{01}\frac{m^2_b}{M\eta_b}+\frac{m_b}{M}X^{(b)}_{02}\right).\label{11}
\end{eqnarray}

From (\ref{eqsda}), (\ref{eqsd1a}), (\ref{21}), (\ref{11}) we can conclude that because of momentum noncommutativity the particles do not move together. In noncommutative phase space the system of free particles with the same initial velocities flies away.

 We would like to mention that in the case when condition on the parameter of noncommutativity (\ref{cond}) is satisfied, from (\ref{eqsda}), (\ref{eqsd1a}) we have that the particles move with the the same velocities which are equal to the velocity of the center-of-mass of the system
\begin{eqnarray}
\dot{X}^{(a)}_1(t)=\sum_a\mu_a\dot{X}^{(a)}_1(t)=\upsilon_{01}\cos\alpha t+\upsilon_{02}\sin\alpha t,\\
\dot{X}^{(a)}_2(t)=\sum_a\mu_a\dot{X}^{(a)}_2(t)=\upsilon_{02}\cos\alpha t-\upsilon_{01}\sin\alpha t.
\end{eqnarray}
 The relative coordinates do not depend on time. Taking into account (\ref{cond}), from (\ref{21}), (\ref{11}) we have
 $\Delta{X}_1^{(a)}=X^{(a)}_{01}-\sum_b\mu_bX^{(b)}_{01}$, $\Delta{X}_2^{(a)}=X^{(a)}_{02}-\sum_b\mu_bX^{(b)}_{02}$.
 Also, the trajectories of free particles do not depend on they masses. Taking into account (\ref{cond}), (\ref{eqs}), (\ref{eqs22}), we have
\begin{eqnarray}
{X}^{(a)}_1(t)=\frac{\upsilon_{01}}{\alpha}\sin\alpha t-\frac{\upsilon_{02}}{\alpha}\cos\alpha t+\frac{\upsilon_{02}}{\alpha}+X^{(a)}_{01},\label{eqs1}\\
{X}^{(a)}_2(t)=\frac{\upsilon_{02}}{\alpha}\sin\alpha t+\frac{\upsilon_{01}}{\alpha}\cos\alpha t-\frac{\upsilon_{01}}{\alpha}+X^{(a)}_{02},\label{eqs221}
\end{eqnarray}
 where $X^{(a)}_{01}=X^{(a)}_1(0)$, $X^{(a)}_{02}=X^{(a)}_2(0)$.
Therefore particles with the same initial velocities at the initial moment of time move together as it should be.

So, in the case when parameter of  noncommutativity  corresponding to a particle depend on its mass as (\ref{cond}) the trajectory and velocity of free particle  in noncommutative phase space do not depend on its mass, the system of free particles with the same initial velocities does not fly away (each particle in the system move with the same velocities which are equal to  velocity of the center-of-mass, the relative coordinates do not depend on time as it should be).

\section{Definition of the momentum of the center-of-mass as integral of motion in noncommutative phase space}
In noncommutative phase space the total momentum (\ref{05}) defined in the traditional way is not the integral of motion. We have
\begin{eqnarray}
\{\tilde{P}_1,H\}=\tilde{\eta}\frac{\tilde{P}_2}{M}+\sum_a\frac{\Delta{P}^{(a)}_2}{m_a}\left(\eta_a-\mu_a\tilde{\eta}\right)\\{}
\{\tilde{P}_2,H\}=-\tilde{\eta}\frac{\tilde{P}_1}{M}-\sum_a\frac{\Delta{P}^{(a)}_1}{m_a}\left(\eta_a-\mu_a\tilde{\eta}\right)
\end{eqnarray}
here $H$ is given by (\ref{hham}).

In the case when conditions on the parameters of noncommutativity (\ref{cond}), (\ref{cond2}) are satisfied we have $\{\tilde{X}_{1},\Delta X_{2}^{(a)}\}=0$, $\{\tilde{P}_1,\Delta{P}^{a}_2\}=0$, therefore
\begin{eqnarray}
\{\tilde{P}_1,H\}=\{\tilde{P}_1,H_{cm}\}=\frac{\tilde{P}_2}{M}\tilde{\eta}\\{}
\{\tilde{P}_2,H\}=\{\tilde{P}_2,H_{cm}\}=-\frac{\tilde{P}_1}{M}\tilde{\eta}.
\end{eqnarray}
here $H_{cm}$ is given by (\ref{12cm}).
 So, because of noncommutativity the total momentum defined as a sum of momenta of particles of the system $\tilde{{\bf P}}=\sum_{a}{\bf P}^{(a)}$ is not preserved in noncommutative phase space.

Let us find the total momentum as integral of motion.
For this purpose let us first consider particular case when a composite system consists of $N$ particles with the same masses $m_1=m_2=...=m_N=m$ and parameters of noncommutativity $\theta_1=\theta_2=...=\theta_N=\theta$, $\eta_1=\eta_2=...=\eta_N=\eta$. Note that in this case coordinates and momenta given by (\ref{05})-(\ref{06}) satisfy $\{\tilde{X}_{1},\Delta X_{2}^{(a)}\}=\{\tilde{P}_1,\Delta{P}^{(a)}_2\}=0$.
We would like to mention that the momenta defined as
\begin{eqnarray}
\tilde{{ P}}_1^{\prime}=\sum_a P^{(a)}_1-\eta \sum_a X^{(a)}_2,\label{pp1}\\
\tilde{{ P}}_2^{\prime}=\sum_a P^{(a)}_2+\eta \sum_a X^{(a)}_1\label{pp2},
\end{eqnarray}
satisfy $\{\tilde{{ P}}_1^{\prime}, H\}=\{\tilde{{ P}}_2^{\prime}, H\}=0$, here $H$ is given by (\ref{hham}). So, they are integrals of motion and can be considered as the total momenta in noncommutative phase space. In the limit $\eta\rightarrow0$ from (\ref{pp1}), (\ref{pp2}) one obtains the total momenta defined in the traditional way.

Let us generalize proposed definition of the total momenta to the case when a composite system consists of $N$ particles with different masses $m_a$ and parameters of noncommutativity $\theta_a$, $\eta_a$.
The total momenta defined as
\begin{eqnarray}
\tilde{{P}}_1^{\prime}=\tilde{P}_1-\tilde{\eta} \tilde{X}_2,\label{prp1}\\
\tilde{{P}}_2^{\prime}=\tilde{P}_2+\tilde{\eta} \tilde{X}_1,\label{prp2}
\end{eqnarray}
with $\tilde{P}_i$, $\tilde{X}_i$, $\tilde{\eta}$ given by (\ref{05}), (\ref{00005}), (\ref{eff2}) are integrals of motion
($\{\tilde{{ P}}_1^{\prime},H\}=\{\tilde{{ P}}_2^{\prime},H\}=0$) in the case when relations (\ref{cond}), (\ref{cond2}) hold.

 Note that for composite system made of $N$ particles with the same masses and parameters of noncommutativity $\eta$, $\theta$ we have $\tilde{X}_i=\sum_a X^{(a)}_i/N$, also, taking into account (\ref{eff2}), we can write $\tilde{\eta}=N\eta$. Therefore from (\ref{prp1}), (\ref{prp2}) we obtain (\ref{pp1}), (\ref{pp2}).

 Let us define the coordinates of the center-of-mass $\tilde{{ X}}_i^{\prime}$ as coordinates conjugated to $\tilde{{ P}}_i^{\prime}$, namely
\begin{eqnarray}
\{\tilde{{X}}_i^{\prime},\tilde{{ P}}_j^{\prime}\}=\delta_{ij},\label{xp}{}
\end{eqnarray}
One can verify that the coordinates defined as
 \begin{eqnarray}
\tilde{{ X}}_i^{\prime}=\frac{\sum_a \mu_a X^{(a)}_i}{1-\tilde{\eta}\tilde{\theta}}=\frac{\tilde{X}_i}{1-\tilde{\eta}\tilde{\theta}},\label{xrp1}
\end{eqnarray}
satisfy (\ref{xp}). Also,  the following relations are satisfied
\begin{eqnarray}
\{\tilde{{X}}_1^{\prime},\tilde{{ X}}_2^{\prime}\}=\frac{\tilde{\theta}}{(1-\tilde{\theta}\tilde{\eta})^2},\label{n1}\\{}
\{\tilde{P}_1^{\prime},\tilde{P}_2^{\prime}\}=\tilde{\eta}(\tilde{\theta}\tilde{\eta}-1).{}\label{n2}
\end{eqnarray}

At the end of this section we would like to mention that the momentum of free particle in noncommutative phase space is not integral of motion (see (\ref{eq1111}), (\ref{eq1})).  Note  that from (\ref{prp1}), (\ref{prp2}) for one-particle system  we can write  $P^{\prime}_1=P_1-\eta X_2$, $P^{\prime}_2=P_2+\eta X_1$ which satisfy $\{P^{\prime}_1,H\}=\{P^{\prime}_2,H\}=0$, where $H$ is given by (\ref{hammm}).  The hamiltonian of the particle in terms of $P_i^{\prime}$ and coordinates $X_i^{\prime}=X_i/(1-\eta\theta)$, is as follows
\begin{eqnarray}
H=\frac{1}{2m}\left(P_1^{\prime}+\eta({1-\eta\theta})X_2^{\prime}\right)^2+\frac{1}{2m}\left(P_2^{\prime}-\eta(1-\eta\theta) X_1^{\prime}\right)^2\label{hhh}
\end{eqnarray}
 We would like to note that the hamiltonian (\ref{hhh}) corresponds to the hamiltonian of a charged particle in the magnetic field ${\bf B}(0,0,B)$ in noncommutative phase space which is characterized by relations  (\ref{xp}), (\ref{n1}), (\ref{n2}), if $eB/c=\eta(1-\eta\theta)$ (here $e$ is charge of the particle,  $c$ is the speed of light).

It  is easy to generalize results presented in the section  to the quantum
case introducing corresponding operators and considering commutation relations for them.

\section{Conclusions}

In the paper we have examined influence of noncommutativity of coordinates and noncommutativity of momenta of canonical type on the motion of composite system. The system of $N$ free particles has been studied. We have found that even for a system of free particles the relative motion in the system affects on the motion of its center-of-mass.
We have shown that because of noncommutativity of momenta the trajectory of free-particle and its velocity depend on mass and parameter of momentum noncommutativity. Therefore free particles of different masses do not move together even in the case when at the initial time the velocities of the particles are the same. So, the system of free particles flies away.

We have concluded that if we consider parameters of noncommutativity which correspond to a particle to be dependent on its mass as  (\ref{cond}) the velocity and trajectory of free particle do not depend on its mass; free particles with the same initial velocities move together, namely each particle in free particles system moves with the velocity which equals to the velocity of the center-of-mass and the trajectory of the relative motion does not depend on time.
We would like to stress that the same condition of the parameters of momentum noncommutativity (\ref{cond}) with condition  on the parameter of coordinate noncommutativity (\ref{cond2}) were proposed in \cite{GnatenkoPLA17}  to solve the problem of violation of the weak equivalence principle, problem of nonadditivity of kinetic energy, problem of dependence of kinetic energy on composition, problem of dependence of motion of the center-of-mass on the relative motion in noncommutative phase space.

We have also shown that in the case when relations (\ref{cond}), (\ref{cond2}) hold the total momenta can be defined as integrals of motion in noncommutative phase space. We have introduced the total  momenta (\ref{prp1}), (\ref{prp2}) and found coordinates of the center-of-mass conjugated to them (\ref{xrp1}). It is shown that in the limit $\tilde{\eta}\rightarrow0$  proposed expressions  for coordinates of the center-of-mass  (\ref{xrp1}) and for the total momenta (\ref{prp1}), (\ref{prp2}) reduce to the ordinary definitions. This is in agreement with the result of paper \cite{GnatenkoPLA13} where it was shown that in a space with noncommutative coordinates ($\theta\neq0$, $\eta=0$) the total momenta defined in the traditional way are integrals of motion. In the limits $\tilde{\theta}\rightarrow0$, $\tilde{\eta}\rightarrow0$ from (\ref{xrp1}), (\ref{prp1}), (\ref{prp2}) one obtains the coordinates and momenta of the center-of-mass defined in the traditional way which satisfy the ordinary Poisson brackets.

So, relations of parameters of noncommutativity which correspond to a particle with its mass (\ref{cond}), (\ref{cond2}) give the possibility to solve number of problems in noncommutative phase space.

\section*{Acknowledgments}
This work was partly supported by the grant of the President of Ukraine for support of scientific researches of young scientists ($\Phi$-75) and by the project $\Phi\Phi$-30$\Phi$ (No. 0116U001539) from the Ministry of Education and Science of Ukraine.


\begin{thebibliography}{9}

\bibitem{Witten} N.~Seiberg, E.~Witten,  {\it J. High Energy Phys.} {\bf 9909}, 032 (1999).
\bibitem{Doplicher} S.~Doplicher, K.~Fredenhagen, J.E.~Roberts,  {\it Phys. Lett. B} {\bf 331}, 39 (1994).
\bibitem{Snyder} H.~Snyder, {\it Phys. Rev.} {\bf 71}, 38 (1947).
\bibitem{Hatzinikitas} A. Hatzinikitas, I. Smyrnakis, {\it J. Math. Phys.} {\bf43},  113 (2002).
\bibitem{Kijanka} A. Kijanka, P. Kosinski, {\it Phys. Rev. D} {\bf70},  127702 (2004).
\bibitem{Jing} Jing Jian, Jian-Feng Chen, {\it Eur. Phys. J. C} {\bf60}, 669 (2009).
\bibitem{Smailagic} A. Smailagic, E. Spallucci, {\it Phys. Rev. D} {\bf65}, 107701 (2002).
\bibitem{Smailagic1} A. Smailagic, E. Spallucci, {\it J. Phys. A} {\bf35}, 363 (2002).
\bibitem{Alvarez} P. D. Alvarez, J. Gomis, K. Kamimura, M. S. Plyushchay, {\it Phys. Lett. B} {\bf 659}  906 (2008).
\bibitem{Djemai} A. E. F. Djemai, H. Smail, {\it Commun. Theor. Phys.} {\bf41}, 837 (2004).
\bibitem{Giri}  P. R. Giri, P. Roy, {\it Eur. Phys. J. C} {\bf57}, 835 (2008).
\bibitem{Geloun} J. Ben Geloun,  S. Gangopadhyay, F. G. Scholtz, {\it EPL} {\bf86}, 51001 (2009).
\bibitem{Nath} D. Nath, P. Roy, {\it Ann. Phys}, {\bf377},  115  (2017).
\bibitem{GnatenkoJPS17}  Kh. P. Gnatenko, V. M. Tkachuk, {\it J. Phys. Stud.} {\bf 21},  3001  (2017).
\bibitem{Shyiko} Kh. P. Gnatenko, O. V. Shyiko,	arXiv:1804.11110.


\bibitem{Gamboa1} J. Gamboa, M. Loewe, F. Mendez, J. C. Rojas, {\it Mod. Phys. Lett. A} {\bf16}, 2075 (2001).
\bibitem{Horvathy} P. A. Horvathy,  {\it Ann. Phys.} {\bf299}  128 (2002).
\bibitem{Dayi} O. F. Dayi, L. T. Kelleyane, {\it Mod. Phys. Lett. A} {\bf17}  1937 (2002).
\bibitem{Horvathy05} P. A. Horvathy, M. S. Plyushchay, {\it Nucl. Phys. B} {\bf 714} 269 (2005).
\bibitem{Alvarez09} P. D. Alvarez, J. L. Cortes, P. A. Horvathy, M. S. Plyushchay, {\it JHEP} {\bf 0903}, 034 (2009).
\bibitem{Daszkiewicz1} M. Daszkiewicz, {\it Acta Phys. Polon. B} {\bf44}, 59 (2013).


\bibitem{Bertolami1} O. Bertolami, J. G. Rosa, C. M. L. de Aragao, P. Castorina, D. Zappala, {\it Phys. Rev. D }{\bf72}, 025010 (2005)
\bibitem{Bastos} C. Bastos, O. Bertolami,  {\it Phys. Lett. A} {\bf372}, 5556 (2008).
\bibitem{GnatenkoMPLA16} Kh. P. Gnatenko, V. M. Tkachuk, {\it Mod. Phys. Lett. A} {\bf31}, 1650026  (2016).
\bibitem{GnatenkoJPS18} Kh. P. Gnatenko, O. O. Morozko, Yu. S. Krynytskyi, {\it J. Phys. Stud.} {\bf22}, 1001 (2018).

\bibitem{Balachandran10} A. P. Balachandran, A. Ibort, G. Marmo, M. Martone, {\it SIGMA} {\bf6}, 052 (2010).
\bibitem{Balachandran11} A. P. Balachandran, A. Ibort, G. Marmo, M. Martone, {\it JHEP} {\bf1103}, 057 (2011).


\bibitem{Ho} Pei-Ming Ho, Hsien-Chung Kao,  {\it Phys. Rev. Lett.} {\bf 88}, 151602 (2002).
\bibitem{Bellucci} S.~Bellucci, A.~Yeranyan,  {\it Phys. Lett. B} {\bf 609}, 418 (2005).
\bibitem{GnatenkoPLA13}  Kh.P.~Gnatenko,  {\it Phys. Lett. A} {\bf 377}, 3061 (2013).
\bibitem{GnatenkoJPS13} Kh. P. Gnatenko, {\it J. Phys. Stud.} {\bf17}, 4001 (2013).

\bibitem{Jabbari} I.~Jabbari, A.~Jahan, Z~ Riazi,  {\it Turk. J. Phys.} {\bf 33}, 149 (2009).
\bibitem{GnatenkoU16} Kh. P. Gnatenko, V. M. Tkachuk, {\it Ukr. J. Phys.} {\bf 61}, 432 (2016).


\bibitem{Daszkiewicz} M.~Daszkiewicz, C.J.~Walczyk,  {\it Mod. Phys. Lett. A} {\bf 26}, 819 (2011).
\bibitem{Daszkiewicz2} M.~Daszkiewicz, {\it Acta Phys. Polon. B} {\bf44} 699 (2013).
\bibitem{GnatenkoPLA17}  Kh. P. Gnatenko, V. M. Tkachuk,  {\it Phys. Lett. A} {\bf 381},  2463 (2017).
\bibitem{GnatenkoIJ18} Kh. P. Gnatenko, V. M. Tkachuk, {\it Int. J. Mod. Phys. A} {\bf 33},  1850037  (2018).
\bibitem{Jacobson} T. Jacobson, S. Liberati, D. Mattingly, {\it Ann. Phys.} {\bf 321}  150 (2006).
\bibitem{Amelino-Camelia} G. Amelino-Camelia, {\it Symmetry} {\bf 2} 230 (2010).
\bibitem{Amelino-Camelia1} G. Amelino-Camelia, L. Freidel, J. Kowalski-Glikman, L. Smolin, {\it Phys. Rev. D} {\bf84}
087702 (2011).
\bibitem{Hossenfelder}  S. Hossenfelder, {\it Phys. Rev. D} {\bf 88}  028701 (2013).
\bibitem{Amelino-Camelia3} G. Amelino-Camelia, L. Freidel, J. Kowalski-Glikman, L. Smolin, {\it Phys. Rev. D} {\bf 88}
028702 (2013).
\bibitem{Quesne} C.~Quesne, V.M.~Tkachuk,  {\it Phys. Rev. A} {\bf 81}, 012106 (2010).
\bibitem{Tkachuk} V.M.~Tkachuk,  {\it Phys. Rev. A} {\bf 86}, 062112 (2012).


\bibitem{GnatenkoMPLA17}  Kh. P. Gnatenko,  {\it Mod. Phys. Lett. A}  {\bf 32}, 1750166 (2017).

\end{thebibliography}
\end{document}